# An Investigation on Thermo-hydraulic Performance of a Flat-plate Channel with Pyramidal Protrusions

Amin Ebrahimi[1, *], Benyamin Naranjani[2]

1- Department of Materials Science & Engineering, Delft University of Technology, Mekelweg 2, 2628 CD Delft, The Netherlands.

2- High Performance Computing (HPC) Laboratory, Department of Mechanical Engineering, Faculty of Engineering, Ferdowsi University of Mashhad, Mashhad, P.O. Box 91775-1111, Khorasan Razavi, Iran.

* - Corresponding Author, Email: A.Ebrahimi@tudelft.nl (A. Ebrahimi), Phone: +31 (0)15 27 85682, Address: Department of Materials Science & Engineering, Delft University of Technology, Mekelweg 2, 2628 CD Delft, The Netherlands.

**Abstract**

In this study, a flat-plate channel configured with pyramidal protrusions are numerically analysed for the first time. Simulations of laminar single-phase fluid flow and heat transfer characteristics are developed using a finite-volume approach under steady-state condition. Pure water is selected as the coolant and its thermo-physical properties are modelled using a set of temperature-dependent functions. Different configurations of the channel, including a plain channel and a channel with nature-inspired protruded surfaces, are studied here for Reynolds numbers ranging from 135 to 1430. The effects of the protrusion shape, size and arrangement on the hydrothermal performance of a flat-plate channel are studied in details. The temperature of the upper and lower surfaces of the channel is kept constant during the simulations. It is observed that utilizing these configurations can boost the heat transfer up to 277.9% and amplify the pressure loss up to 179.4% with a respect to the plain channel. It is found that the overall efficiency of the channels with pyramidal protrusions is improved by 12.0%-169.4% compared to the plain channel for the conditions studied here. Furthermore, the thermodynamic performance of the channel is investigated in terms of entropy generation and it is found that equipping the channels with pyramidal protrusions leads to lower irreversibility in the system.

# 1. Introduction

Compact heat exchangers are very common in different engineering applications such as automotive and aerospace industries, heating and refrigerating, solar collectors, electronic devices, laser technology. In recent decades lots of efforts have been made to improve thermal performance of the compact heat exchangers accompanying a reduction in their size, weight and cost. The heat transfer can be boosted using active and/or passive techniques [1, 2]. A variety of passive techniques such as flow additives, swirl flow devices, surface tension devices, rough surfaces, treated surfaces, pin fins, ribbed turbulators and surfaces with dimple and/or protrusions are used for enhancing heat transfer in different applications. The performance of these techniques for enhancing the heat transfer rates are compared to each other by Ligrani et al. [3].

Protruded surfaces are classified as one of the passive heat transfer enhancement methods and can significantly enhance the heat transfer by reducing the thermal resistance of the sublayer adjacent to the solid walls. This is done by generating secondary flows, disrupting the boundary layer growth, flow recirculation and shear-layer reattachment, promoting mixing and increasing the turbulence intensity [4]. In the other hand, using protruded surfaces in thermal systems causes a higher pressure drop due to the losses induced by secondary flow, increasing shear-stresses and velocity gradients, and intensive interactions between vortices and the channel walls [5]. Hwang et al. [6] experimentally studied the heat transfer performance of different protrusion/dimple patterned surfaces within a rectangular channel. They reported that for a case with double-side patterned surfaces the overall heat transfer coefficient is much greater than that of a single-side patterned surface thanks to stronger mixing flow. Chen et al. [7] numerically investigated hydro-thermal characteristics of a turbulent channel flow with densely arranged protrusions on its walls. They observed that the higher the height of the protrusions the higher the heat transfer and friction factor. They found an extremum in performance factor curve with increasing the height of the protrusions.

One can find that most of the literature have focused on evaluating the impacts of the hemispherical protrusions on heat transfer characteristics of turbulent channel flows [8-10]; whereas the investigations on flow structure and heat transfer characteristics of protrusions with different shapes inside the channels, especially under laminar flow condition, are scarce. It is well known that the conventional hemispherical protrusions are no longer worthy for increasing demands of heat removal applications; therefore, the researchers are moving towards novel structures and combining different techniques to design more efficient systems in recent years [11-16]. The main objective of this paper is introducing a novel protruded surface to enhance the heat transfer performance of heat exchangers. In order to achieve this goal a novel surface pattern is designed which is inspired from the skin patterns of the desert plants and animals such as cactuses, alligators and thorny dragons. According to the best of the authors' knowledge, it is the first time that pyramidal protrusions are employed for heat transfer augmentation purposes. The effects of utilizing pyramidal protrusions on the laminar flow pattern and heat transfer performance are scrutinized in this paper. Different configurations of the flat-plate channel with pyramidal protrusions, including various alignments (inline and staggered), angle of attacks and sizes, are investigated. Furthermore, the thermodynamic performance of the channel is studied using entropy generation analysis.

**2. Model Descriptions**

**2.1. Geometric configurations and computational domain**

In this paper, three-dimensional simulations are carried out on different configurations of a flat-plate channel with and without obstacles. Obstacles in the form of protrusions are mounted on both the top and bottom walls of the channel. The schematic diagram of the computational domain and relevant geometrical parameters are illustrated in figure 1. The height (H) of the channel is parametrized with the width of the channel (W) and is 3W/4. The computational domain consists of three zones, namely, inlet zone, main zone and outlet zone. The inlet zone is considered at the entrance of the main zone to ensure the flow uniformity before the protrusions. Furthermore, the outlet zone is embedded after the main zone to ensure that there is no back flow at the outlet boundary. The length of the inlet zone ($L_i$) and the outlet zone ($L_o$) are selected to be half of the length of the main zone (L=20W) [17-19]. Nineteen equally spaced pyramidal protrusions ($L_b$=W) are located in the main zone with inline and staggered arrangements. It is worth mentioning that the minimum distance between the main zone entrance and the centroid of the pyramid's base at the first row equals the channel width (W). Protrusions in the form of a square-based right pyramids are defined by the base edge length (a) and apex height ($H_v$) with different aspect ratios (AR=a/$H_v$). The flow is described in a three-dimensional Cartesian coordinate system in which x is the span-wise direction, y is the normal direction and z is the stream-wise direction. It should be noted that the origin of the z axis is located at the entrance of the main zone.

## 2.2. Mathematical Methods, Governing Equations and boundary conditions

Simulations are performed to scrutinize flow pattern and heat transfer characteristics inside a flat-plate channel with protruded surfaces. Pure water is chosen to be the coolant and its thermo-physical properties are modelled using a set of temperature-dependent functions as summarised in Table 1. In table 1, $\rho$, $k$ and $\mu$ are density, thermal conductivity and dynamic viscosity of the fluid, respectively. The flow is assumed to be incompressible, Newtonian and laminar due to low fluid velocity and the mild incidence angle between the flow and protrusions. Moreover, radiation effects and body forces are assumed to be neglected in this study. Therefore the conservative, steady-state form of continuity, momentum and energy equations can be expressed, respectively as below:

$$\nabla \cdot \vec{V} = 0. \tag{1}$$

$$\rho \vec{V} \cdot \nabla \vec{V} = -\nabla p + \nabla(\mu \nabla \vec{V}) \tag{2}$$

$$\rho c_p (\vec{V} \cdot \nabla T) = \nabla(k \nabla T) + \Phi \tag{3}$$

where, $\vec{V}$ is velocity vector, $\rho$ is density, $p$ is static pressure, $\mu$ is dynamic viscosity, $c_p$ is specific heat capacity, $T$ is temperature, $k$ is thermal conductivity and $\Phi$ is related to dissipation function which can be extended as below.

$$\Phi = \left[\left(\mu \frac{\partial u}{\partial x}\right)^2 + \left(\mu \frac{\partial v}{\partial y}\right)^2 + \left(\mu \frac{\partial w}{\partial z}\right)^2\right] + \\ \left[\left(\mu \frac{\partial u}{\partial y} + \mu \frac{\partial v}{\partial x}\right)^2 + \left(\mu \frac{\partial v}{\partial z} + \mu \frac{\partial w}{\partial y}\right)^2 + \left(\mu \frac{\partial w}{\partial x} + \mu \frac{\partial u}{\partial z}\right)^2\right] \tag{4}$$

The required boundary conditions for conducting the numerical simulations can be introduced as follow.

Inlet boundary (1-2-3-4):

$$u = v = 0., w = U_{in} = cte., T = T_{in} = 298.15\,(K) \tag{5}$$

Heated walls which contain (5-6-10-9, 8-7-11-12) and faces of the pyramid:

$$u = v = w = 0, T = T_{wall} = 348.15\,(K) \tag{6}$$

Symmetry boundaries (1-13-16-4, 2-14-15-3):

$$u = 0., \frac{\partial v}{\partial x} = \frac{\partial w}{\partial x} = \frac{\partial T}{\partial x} = 0. \tag{7}$$

Adiabatic walls (1-2-6-5), (4-3-7-8), (9-10-14-13) and (12-11-15-16):

$$u = v = w = 0., \frac{\partial T}{\partial y} = 0. \tag{8}$$

Outlet (13-14-15-16):

$$\frac{\partial u}{\partial z} = \frac{\partial v}{\partial z} = \frac{\partial w}{\partial z} = 0., \frac{\partial T}{\partial z} = 0. \tag{9}$$

### 2.3. Numerical procedures and parameter definitions

In this work, an open-source computational fluid dynamic package (OpenFOAM v3.0) is utilized to solve the governing equations. The discretization of the computational domain is done by non-uniform structured hexahedral grids. In order to achieve a high quality mesh and have a better control on grid sizes, the computational domain is divided into a number of simple zones. The grids are well refined near the walls and around the protrusions.

The aforementioned governing equations are discretized by the finite-volume approach. The SimpleFOAM flow solver is used as starting point and is extended to temperature equation. The SIMPLEC method is used for pressure-velocity coupling [20]. The upwind scheme is utilized for the discretization of the convection term and the central difference scheme is employed for the discretization of the diffusion term both with second order accuracy. The equations are solved iteratively with an implicit scheme based on a pressure-based solver. The iterative process of solving the governing equations maintained until the residuals of the continuity and the momentum equations become less than $10^{-6}$ and for the energy equation the residual value become less than $10^{-8}$. The following parameters are defined to represent the results of the present numerical simulations. The Reynolds number ($Re$) is defined as a function of the channel hydraulic diameter ($D_h$) as follow.

$$Re = \frac{\rho U_{in} D_h}{\mu} \tag{10}$$

where $D_h$ is the distance between the upper and lower surfaces of the channel.

The apparent friction factor (*f*) can be calculated as follow.

$$f = \frac{2\Delta p}{\rho U_{in}^2} \frac{D_h}{L} \tag{11}$$

$$\Delta p = \left(\bar{p}_o - \bar{p}_i\right) \tag{12}$$

$$\bar{p}_z = \frac{\int p \, dA}{\int dA} \tag{13}$$

where $\Delta p$ is the pressure drop through the main zone, L is the length of the main zone and $\bar{p}_z$ is the area-weighted average of the cross sectional static pressure. The subscripts of *o* and *i* stand for the outlet and inlet cross sections of the main zone, respectively.

The heat flux (q) and mean Nusselt number (*Nu*) can be defined by subsequent relations.

$$Nu = \frac{D_h}{k_f} \ln\left(\frac{T_{wall} - \bar{T}_{i,m}}{T_{wall} - \bar{T}_{o,m}}\right) \frac{\dot{m} c_{p,f}}{A_{ht}} \tag{14}$$

$$q = \frac{Q}{A_{ht}} \tag{15}$$

$$Q = \dot{m} c_p \left(\bar{T}_{o,m} - \bar{T}_{i,m}\right) \tag{16}$$

In the aforementioned equations, $\dot{m}$ is the mass flow rate through the channel, $A_{ht}$ is the total heat transfer area, $c_{p,f}$ and $k_f$ are specific heat capacity and thermal conductivity of the fluid at the arithmetic mean temperature of the outlet and inlet. Additionally, *Q* is the total heat transfer rate, $T_{wall}$ is the temperature of the heated walls, $\bar{T}_{i,m}$ and $\bar{T}_{o,m}$ are mass-weighted average temperatures of outlet and inlet cross sections at the inlet and outlet of the channel zone, respectively.

To assess thermodynamic performance of the channels with protrusions, the total volumetric entropy generation rate ($\dot{S}_g'''$), based on the obtained velocity and temperature distribution across the computational domain can be calculated as follow [21, 22].

$$\dot{S}_g''' = \dot{S}_{g,\Delta p}''' + \dot{S}_{g,\Delta T}''' \tag{17}$$

$$\dot{S}_{g,\Delta p}''' = \frac{\mu}{T}\left\{ 2\left[\left(\frac{\partial u}{\partial x}\right)^2 + \left(\frac{\partial v}{\partial y}\right)^2 + \left(\frac{\partial w}{\partial z}\right)^2\right] + \left(\frac{\partial u}{\partial y} + \frac{\partial v}{\partial x}\right)^2 + \left(\frac{\partial u}{\partial z} + \frac{\partial w}{\partial x}\right)^2 + \left(\frac{\partial v}{\partial z} + \frac{\partial w}{\partial y}\right)^2 \right\} \tag{18}$$

$$\dot{S}_{g,\Delta T}''' = \frac{k}{T^2}\left[\left(\frac{\partial T}{\partial x}\right)^2 + \left(\frac{\partial T}{\partial y}\right)^2 + \left(\frac{\partial T}{\partial z}\right)^2\right] \tag{19}$$

Where, $\dot{S}_{g,\Delta p}'''$ and $\dot{S}_{g,\Delta T}'''$ are entropy generations due to flow friction and heat transfer, respectively.

The total entropy generation rate can be non-dimensionalised into ($S_{G,total}$) as follows [23].

$$S_{G,total} = \dot{S}_g''' \frac{k T_{in}^2}{q^2} \tag{20}$$

## 3. Grid independency and model verification

In order to attain a reliable and accurate solution independent from the grid size with a reasonable computational cost, five different grids are checked for the grid independence test. H1 configuration is considered with grid sizes varying from 787,400 (very coarse) to 1,709,900 (very fine). The results of apparent friction factor (*f*) and mean Nusselt number (*Nu*) is reported in Table 2 for *Re*=715. The deviation of the former parameters for the meshes with fine and very fine grids are well below 1%, hence, a mesh with fine grids, including 1,393,100 grids is selected to conduct the computations which guarantees appropriate precision of the results.

In order to ensure that the present numerical model can predict acceptable results, simulations are carried out to compare the obtained results with available experimental and numerical data reported in [17]. The simulations are done for water flow inside a channel with rectangular, triangular and trapezoidal vortex generators with the same boundary conditions. One can find more about the test cases and the experimental implementations in reference [17]. As reported in table 3, the results of the present numerical model show a good agreement with the reported experimental and numerical data. It is clear that the maximum deviation of the present numerical results from the experimental data is lower than 6.2% and 2.5% for pressure drop and heat transfer coefficient, respectively. Additionally, the maximum difference between the results of the proposed model and numerical data of [17] is lower than 5.6% and 4.7% for heat transfer coefficient and pressure drop, respectively. It is worth mentioning that the authors of [17] reported that their numerical model under-predicts the experimental pressure drop by 8.9% and over-predicts the heat transfer coefficient obtained from experiments by 4.6%. Taking into account the experimental uncertainties and model simplifications, the obtained results are satisfactory.

## 4. Results and discussions

In the present study, simulations of laminar single-phase fluid flow and heat transfer are performed under steady-state condition for various configurations of the channel with protruded surfaces for Reynolds numbers ranging from 135 to 1430. Additionally, a plain channel with smooth surfaces is considered to evaluate the heat transfer enhancement and overall thermal efficiency of the designed configurations. It should be noted that the base edge length (a) and apex height ($H_v$) of the pyramidal protrusions is W/2 and W/4, respectively, through this paper unless stated.

Figure 2 depicts the variation of the apparent friction factor as a function of Reynolds number. It is clearly seen that the lowest friction factor for each value of the Re belongs to the plain channel. It is argued that strong interactions of the generated vortices with each other and with the walls, as well as main flow acceleration with decreasing the cross-section main flow area is the reason of higher pressure drop for the channels with protrusions. Moreover, more pressure loss is brought in the regions that the protrusions are placed. It is seen that the H1 and H2 configurations show higher friction factors. It is worth mentioning that for both H1 and H2 configurations the protrusions are mounted with α=45°. For the H1 and H2 configurations the cross sectional flow area is smaller than that of the other configurations which causes more flow acceleration, higher velocity gradients, higher shear-stress and eventually higher pressure loss. Furthermore, a larger area of the flow domain is affected by the vortices for H1 and H2 channels. It is also observed that *f* will decline with an increase in *Re*.

Variations of mean Nusselt number ($Nu_m$) with *Re* are presented in Figure 3 for different configurations. It is clearly seen that mounting pyramidal protrusions could enhance the heat transfer in the channel. It is found that the proposed configurations can boost the heat transfer by 37.8% - 277.9% compared to the plain channel for the range of parameters studied here. H1 and H2 channels in either inline or staggered alignment of the pyramids possess the highest values of *Nu*. It can be explained that for these configurations the generated vortices are mainly longitudinal while transverse vortices are generated inside the H3 and H4 channels. Previous studies on the heat transfer enhancement with vortex generators (VGs) have shown that longitudinal vortices are more effective in heat transfer augmentation [24, 25]. Higher flow acceleration and more intensified flow circulations inside the H1 and H2 channels are other reasons for better heat transfer performance.

Contours of temperature and secondary flow vectors are shown in Figure 4 for two cross sections located at the middle (z/L=0.5) and the end (z/L=1.0) of the main zone for *Re*=1430. Some contra-rotating vortices are observed behind the pyramidal vortex generators which their strength decreases by moving toward the channel outlet. The pyramidal protrusions are responsible for generating the needed pressure gradient to build up the secondary flow vortices. These vortices play the most important role in the heat transfer augmentation of the designed patterns compared to a channel with smooth walls. They transfer the hot fluid near the walls directly in the cold region in the middle of the channel and vice versa. These fluid motions causes stronger mixing flow, disrupt the boundary layer development and makes eddies to penetrate deeper into the sublayer adjacent to the solid surfaces. Considering secondary flow vectors, it is found that for a specified alignment, stronger vortices and more intensified secondary flow are induced for surface patterns in which the protrusions are mounted with $\alpha=45°$. The secondary flow structures are illustrated in Figure 5 for different configurations of the channel. There are generally two counter-rotating pairs of vortices in each stream-wise cross section behind the VGs. For H1 and H3 configurations (i.e. protrusions in an inline alignment), the generated vortices induce an up-wash flow and a down-wash flow between the vortices in the lower region and the upper region of the channel, respectively. It is not the case for H2 and H4 configurations that the protrusions are mounted in a staggered configuration; for these configurations a down-wash flow and an up-wash flow is seen between the vortices in the lower region and the upper region of the channel, respectively. These fluid motions will disrupt the boundary layer development and enlarge the temperature gradient near the walls causing higher heat transfer rates.

The proposed surface patterns not only enhance the heat transfer performance, but also cause higher pressure penalty in the system. In order to assess the effects of proposed surface patterns on the overall efficiency ($\eta_T$) of the flat-plate channels the following parameter is considered [19, 26-30].

$$\eta_T = \frac{Nu/Nu_s}{\left(f/f_s\right)^{1/3}} \tag{21}$$

Figure 6 indicates the variations of $\eta_T$ as a function of Reynolds numbers. It is found that the overall performance of the channels equipped with pyramidal protruded surfaces is higher than that of a plain channel. It is clearly seen that for the channels with protrusions the $\eta_T$ increases with increasing *Re*. It is more reasonable to use the proposed technique for high flow rates under laminar flow regime. Among all the configurations studied in this paper, H1 and H2 have the highest overall efficiencies.

In order to examine the impact of protrusion size on the hydrothermal performance of the channel, various protrusions with different aspect ratios are studied. For this purpose, the H1 configuration is considered because of its high overall performance. Figure 7 shows the variations of $Nu_m$ and *f* with Reynolds number for different base edge lengths and apex heights of the pyramidal protrusions. Strength and location of the generated vortices are one of the most important factors in heat transfer enhancement using protruded surfaces. It is observed that increasing 'a' and/or '$H_v$' results in higher heat transfer rates due to more intensified flow circulation, stronger mixing flow and disrupting the growth of the thermal boundary layer. On the other hand, these effects cause a higher pressure drop and bring more form drag. The effects of pyramidal protrusion height and width variations on overall heat transfer performance are presented in figure 9. It is clearly seen that all the cases with protruded surfaces have a better overall performance compared to the plain channel for the entire range of the Reynolds numbers studied in this paper. The results demonstrate that the larger the protrusion height and width, the higher the overall performance. Furthermore, the overall performance of the channel boosts with increasing the Reynolds number.

A comparison is made between the hydrothermal performance of the proposed surface pattern and that of the common surface patterns for heat transfer enhancement applications. The H1 configuration with pyramidal protrusions with AR=2 (a=W/2 and $H_v$=W/4) is selected as a reference case for this purpose. All cases are designed to have a same total heat transfer area and the obstacles are mounted in a same arrangement as H1. Additionally, the base area of the obstacles is same. Figure 9 illustrates the variations of $Nu_m$ and $f$ as a function of Reynolds number for different obstacle shapes. It is clearly seen that all the surface patterns have higher values of the Nusselt number compared to a plain channel. It can be found that pyramidal protrusions show the highest values of the Nusselt number among all the cases considered here. According to figure 9, despite having better heat transfer performance, using pyramidal protrusions causes higher pressure penalty in the system. Figure 10 shows the overall performance of the channel with different surface patterns. It is seen that the channel with pyramidal protrusions has the highest overall performance thanks to its better heat transfer performance. It is also found that the overall performance of the channels with considered surface patterns increases with Reynolds number.

Systems with a higher degree of irreversibility (or higher entropy generation rate) waste the profitable power and suffer from low efficiency. Minimizing the entropy generation rate will result in higher energy efficiency and therefore lower rates of the entropy generation are desirable [31]. The dimensionless total entropy generation is plotted in Figure 9 for different Reynolds numbers and configurations. It is observed that using pyramidal protrusions inside the flat-plate channels leads to lower rates of entropy generation. It can be seen that total entropy generation rate tends to decline with increasing the *Re*. According to the aforementioned discussions, the diffusive heat transfer is dominant at lower Reynolds numbers and the bulk flow temperature is higher which causes lower fluid viscosity and hence higher temperature and velocity gradients. A minimum in the entropy generation curve does not exist for the ranges of the parameters investigated in this work. It is also found that the contribution of the heat transfer is much more than that of the flow friction in the total entropy generation.

## 5. Conclusions

The flow and heat transfer characteristics of pyramidal protrusions with different configurations were investigated for the first time. A numerical approach using finite-volume method was utilized to study the thermo-hydraulic performance of a channel with pyramidal protrusions in the framework of OpenFOAM. The results were compared with a flat-plate channel with smooth surfaces and protruded surfaces with various protrusion shapes. The accuracy and reliability of the results were confirmed after doing a grid independence test and code validation. The results of the present numerical study were certified by the available experimental and numerical data and the following conclusions were obtained.

Higher heat transfer was observed for the channels with pyramidal protrusions compared to the channel with smooth surfaces due to stronger mixing flow and secondary flow, thinner thermal boundary layer and larger heat transfer surface area. In spite of having better heat transfer performance the application of the designed channels requires more pumping power. From the view of energy savings, all the mentioned configurations with protruded surfaces have better performance compared to a plain channel. The overall performance of the channels with protruded surfaces remarkably improves by increasing the Reynolds number under the laminar flow regime. Among all the configurations studied in this paper, the surface patterns with pyramids mounted at $\alpha=45°$ generate stronger vortices and show the best overall efficiencies. Furthermore, the results demonstrate that the larger the protrusion height and width, the higher the overall performance. According to the second law analysis, the proposed surface pattern is a good option for heat transfer applications and is recommended for novel designs of compact heat exchangers due to lower irreversibility and better thermodynamic performance.

**NOMENCLATURE**

| | |
|---|---|
| a | Base edge length of pyramids, m |
| AR | Protrusion aspect ratio |
| $c_p$ | Specific heat capacity, J/kg.K |
| $D_h$ | Hydraulic diameter, m |
| $f$ | Apparent friction factor |
| H | Height, m |
| $h$ | Convection heat transfer coefficient, W/m$^2$K |
| $k$ | Thermal conductivity, W/mK |
| L | Length, m |
| $Nu$ | Nusselt number |
| $p$ | Pressure, Pa |
| $Q$ | Heat transfer rate, W |
| $q$ | Heat flux, W/m$^2$ |
| $Re$ | Reynolds number |
| S | Distance between apex of pyramids and side walls, m |
| $S_{G,\,tot}$ | Non-dimensional total entropy generation |
| $T$ | Temperature, K |

| | | |
|---|---|---|
| *U* | | Fluid velocity, m/s |
| *u*, *v*, *w* | | Velocity vector components |
| *V* | | Total volume of the heated zone, $m^3$ |
| W | | Width, m |
| x, y, z | | Cartezian coordinates |
| $\dot{m}$ | | Mass flow rate, kg/ $m^3$ |

Greek Symbols

| | |
|---|---|
| α | Attack angle |
| η | Efficiency |
| *μ* | Dynamic viscosity, kg/m.sec |
| *ρ* | Fluid density, $kg/m^3$ |

Subscripts

| | |
|---|---|
| b | Between pyramids |
| f | Fluid |
| ht | Heat transfer |
| i | Inlet |
| m | Mean |
| n | Pyramid's numbers |
| o | Outlet |

s      Simple channel

w      Water

**List of Figures**

Figure 1- Physical model and relevant geometrical parameters of the channel with pyramidal protrusions. (a) 3D perspective view of the computational domain. Top view of (b) H3, (c) H1, (d) H2 and (e) H4 configurations.

Figure 2- Variations of apparent friction factor versus Reynolds number for different configurations.

Figure 3- Effects of different configurations on mean Nusselt number for different Reynolds numbers.

Figure 4- Contours of temperature and secondary flow vectors for different configurations at cross sections located at middle (z/L=0.5; bottom row) and end (z/L=1.0; top row) of the main zone. (a) H1 (b) H2 (c) H3 (d) H4.

Figure 5- Secondary flow structures for different configurations (a) H1 (b) H3 (c) H2 (d) H4.

Figure 6- Variations of overall efficiency as a function of Reynolds number for different configurations.

Figure 7- Effects of pyramidal protrusion height and width variations on (a) apparent friction factor, and (b) mean Nusselt number. (H1 configuration)

Figure 8- Variations of overall efficiency as a function of Reynolds number for different protrusion aspect ratios. (H1 configuration)

Figure 9- Effects of protrusion shape on mean Nusselt number and apparent friction factor for different Reynolds numbers. (H1 configuration)

Figure 10- Variations of overall efficiency as a function of Reynolds number for different protrusion shapes. (H1 configuration)

Figure 11- Dimensionless total entropy generation versus Reynolds number for different configurations.

**List of tables**

Table 1- Thermo-physical properties of pure water [16].

Table 2- The results of grid independence test for H1 configuration at $Re$ =715.

Table 3- Comparison of pressure drop and heat transfer coefficient between numerical results and available experimental and numerical data.

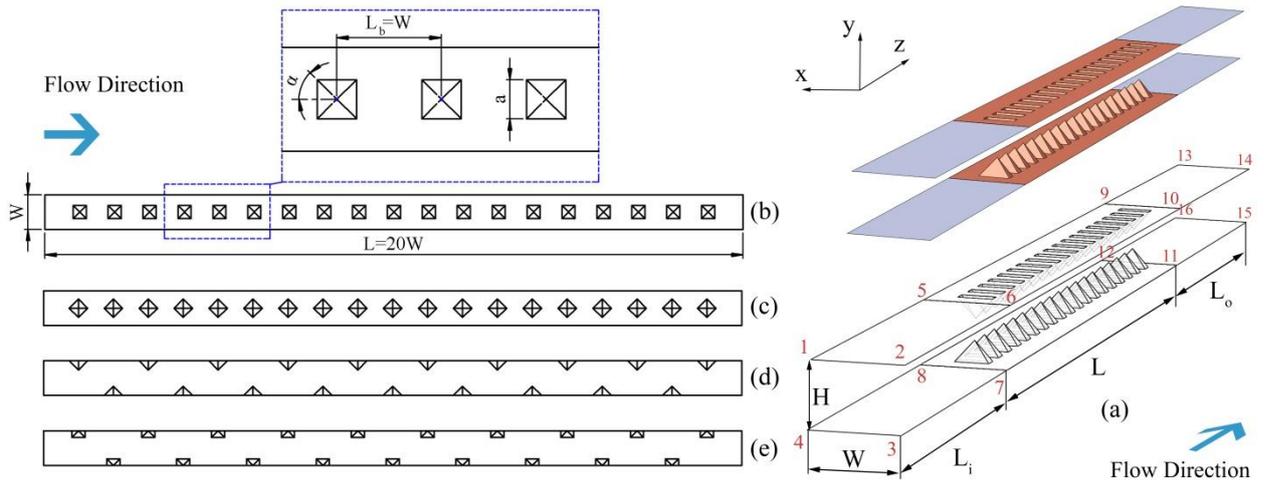

Figure 1- Physical model and relevant geometrical parameters of the channel with pyramidal protrusions. (a) 3D perspective view of the computational domain. Top view of (b) H3, (c) H1, (d) H2 and (e) H4 configurations.

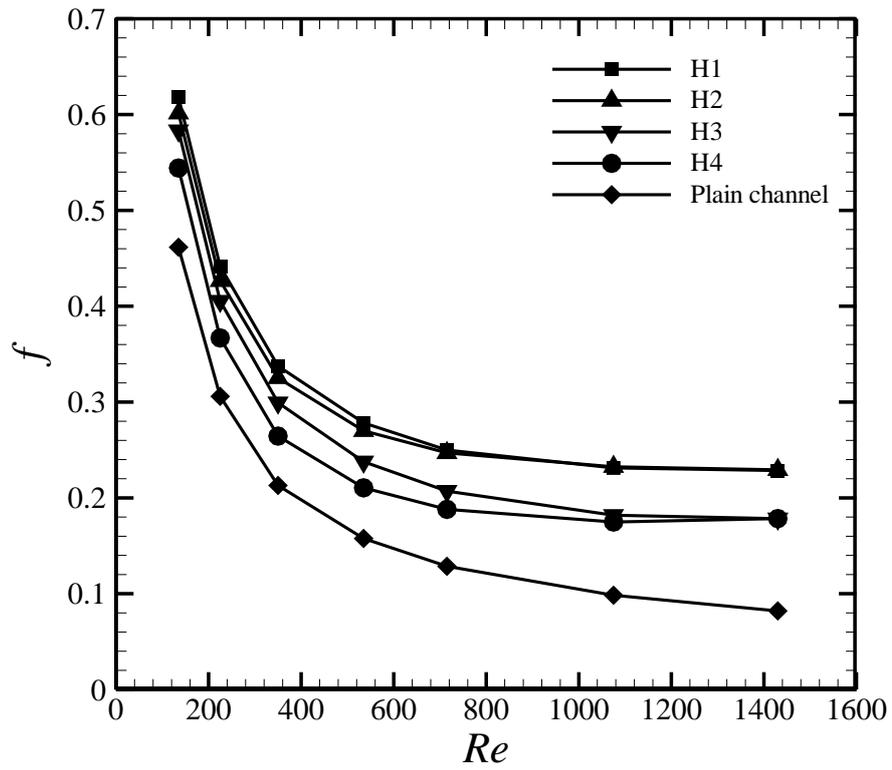

Figure 2- Variations of apparent friction factor versus Reynolds number for different configurations. (a=w/2; $H_v$=w/4)

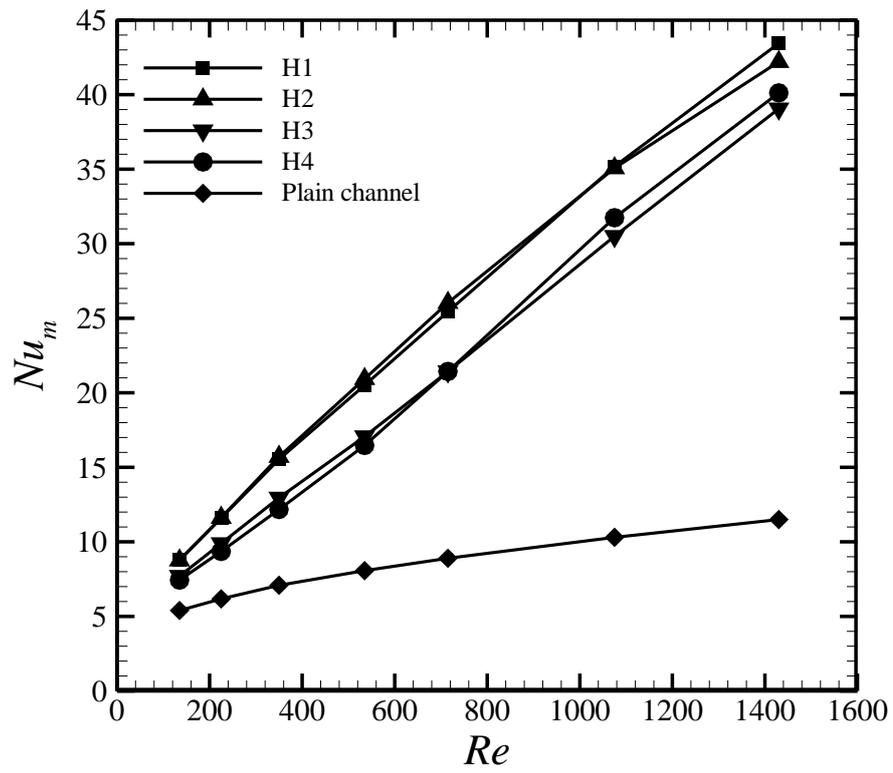

Figure 3- Effects of different configurations on mean Nusselt number for different Reynolds numbers. (a=w/2; $H_v$=w/4)

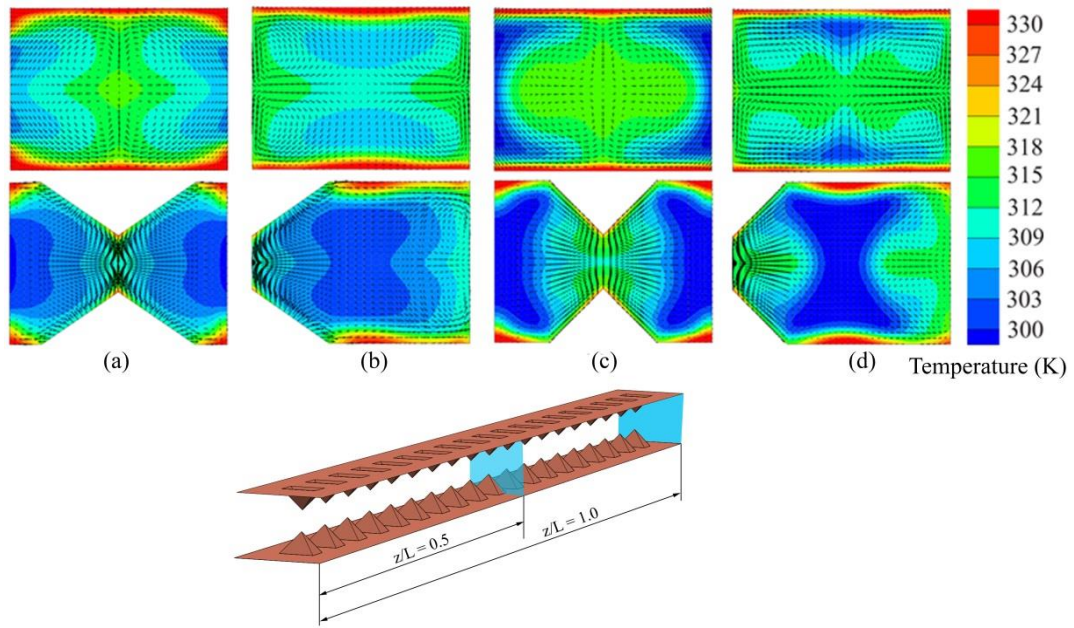

Figure 4- Contours of temperature and secondary flow vectors for different configurations at cross sections located at middle (z/L=0.5; bottom row) and end (z/L=1.0; top row) of the main zone. (a) H1 (b) H2 (c) H3 (d) H4. (a=w/2; $H_v$=w/4)

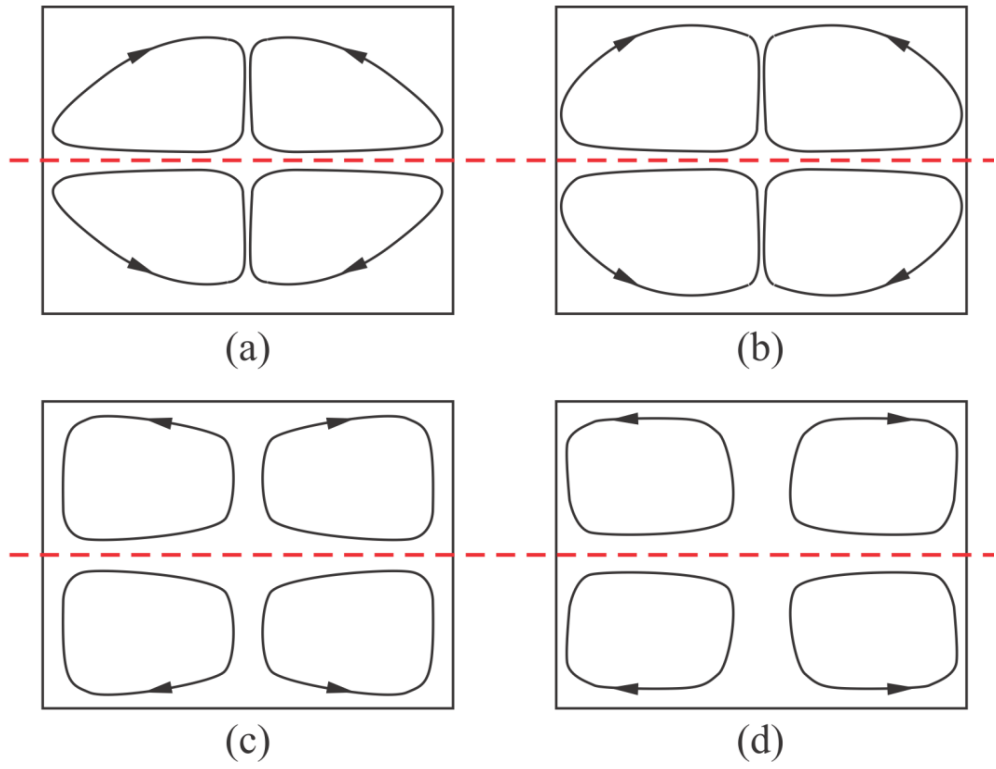

Figure 5- Secondary flow structures for different configurations (a) H1 (b) H3 (c) H2 (d) H4.

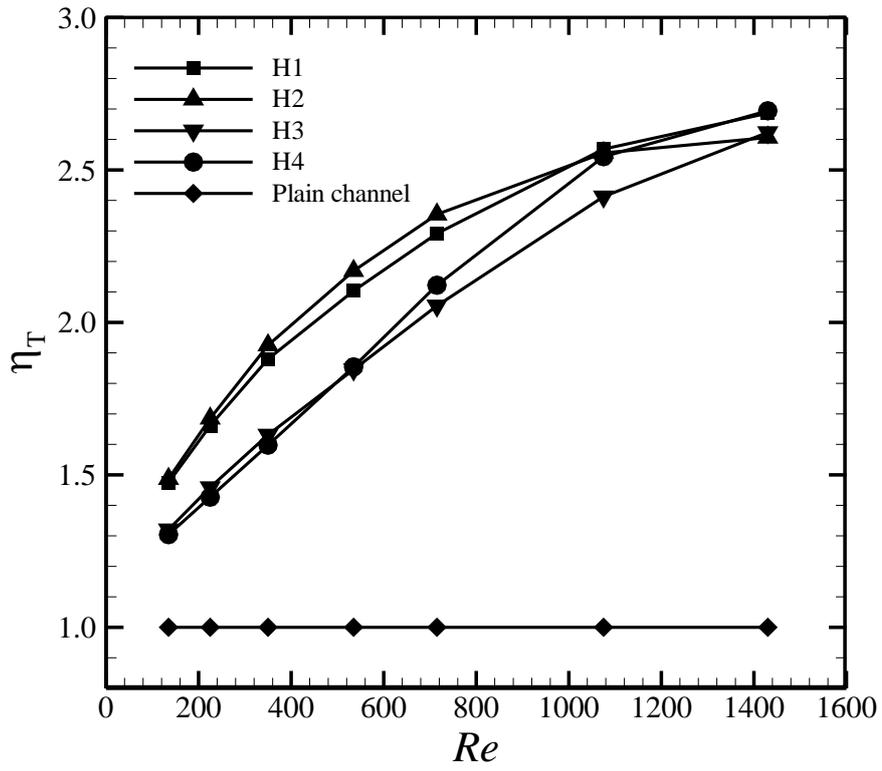

Figure 6- Variations of overall efficiency as a function of Reynolds number for different configurations. (a=w/2; $H_v$=w/4)

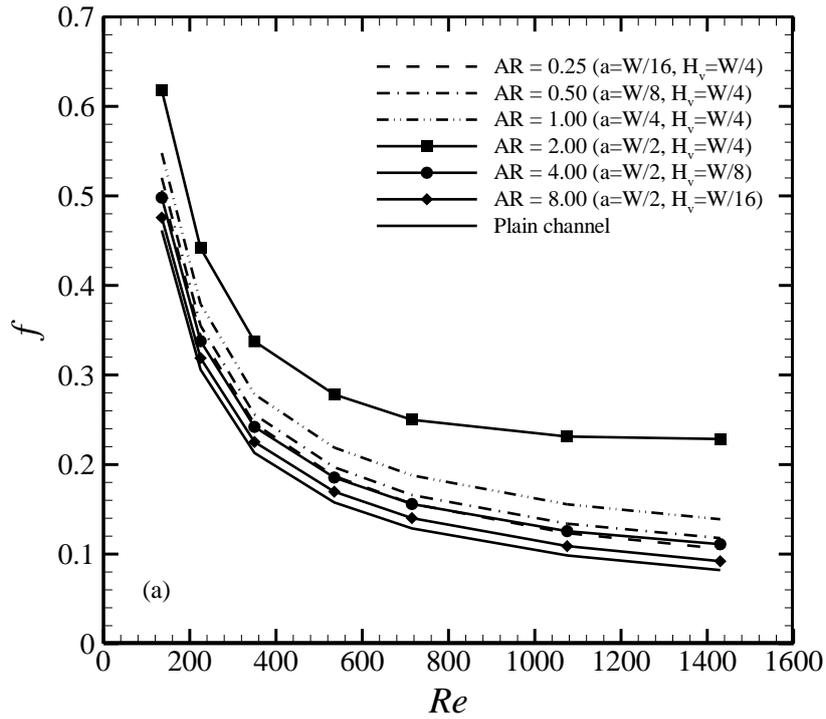

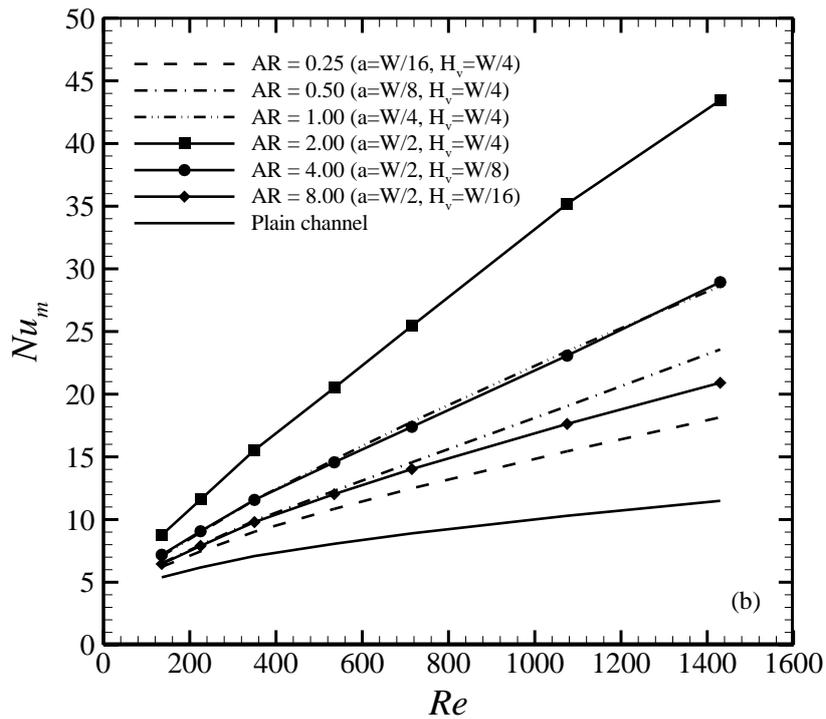

Figure 7- Effects of pyramidal protrusion height and width variations on (a) apparent friction factor, and (b) mean Nusselt number. (H1 configuration)

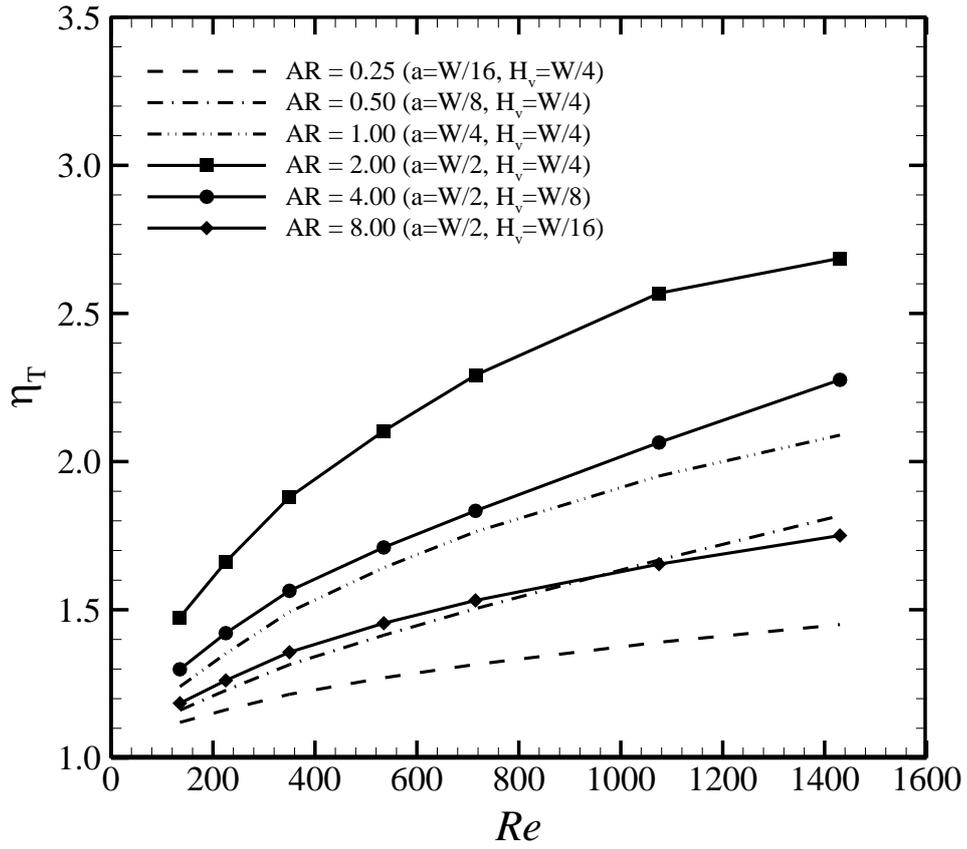

Figure 8- Variations of overall efficiency as a function of Reynolds number for different protrusion aspect ratios. (H1 configuration)

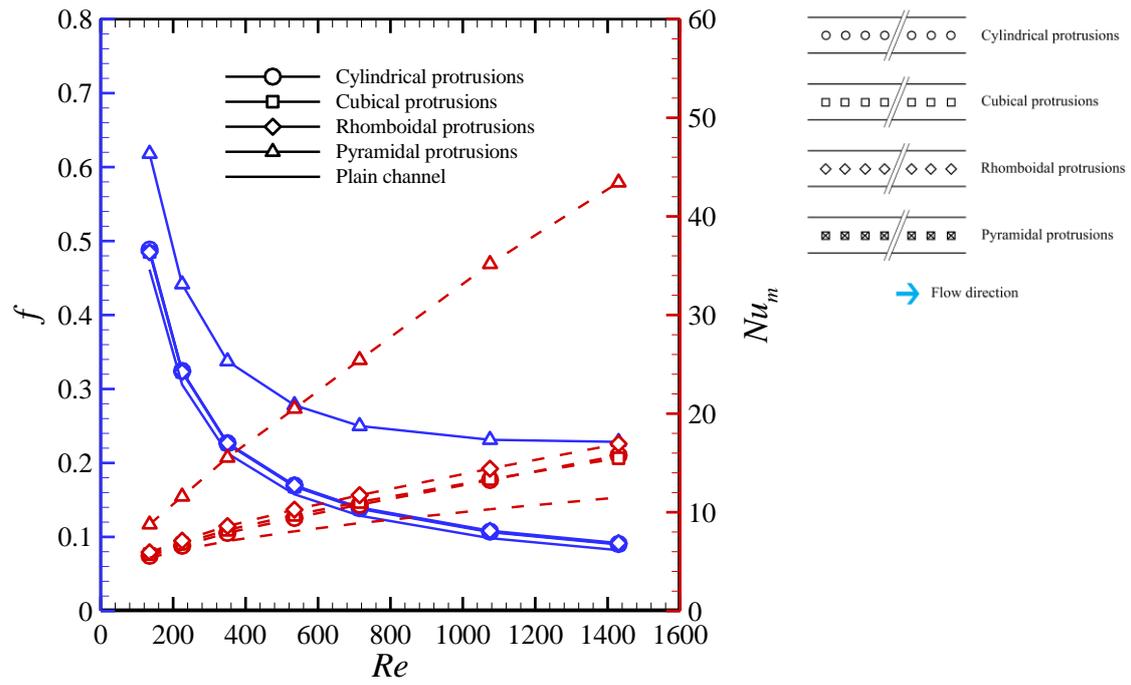

Figure 9- Effects of protrusion shape on mean Nusselt number (red dashed lines) and apparent friction factor (blue solid lines) for different Reynolds numbers. (H1 configuration)

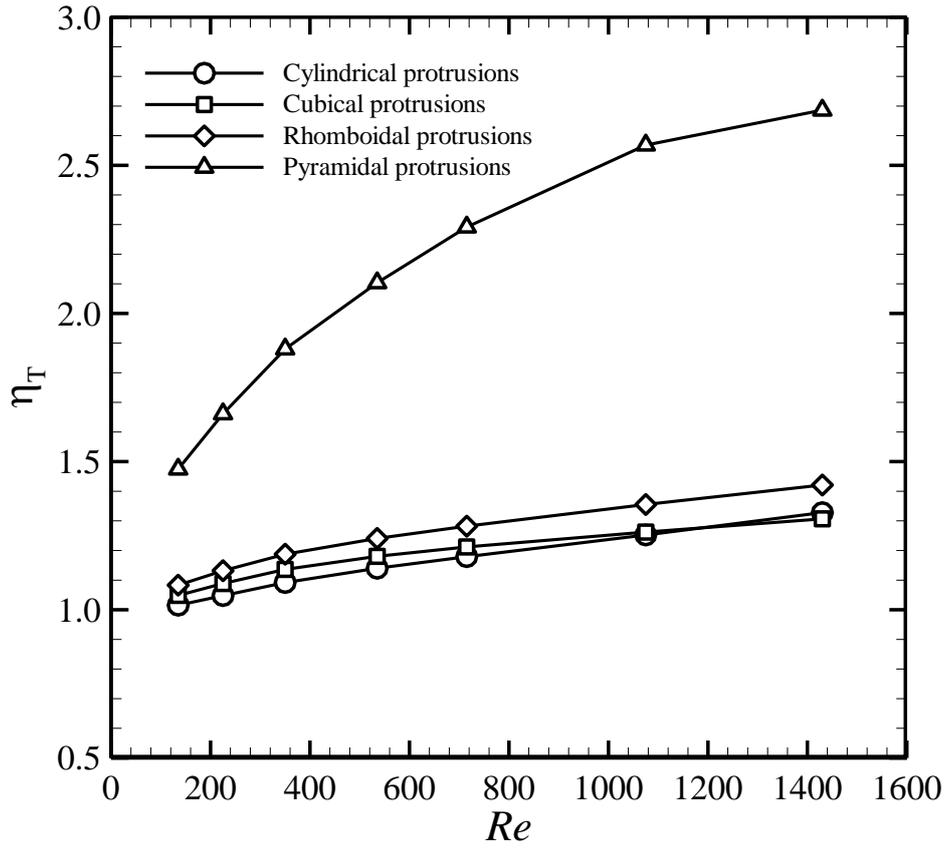

Figure 10- Variations of overall efficiency as a function of Reynolds number for different protrusion shapes. (H1 configuration)

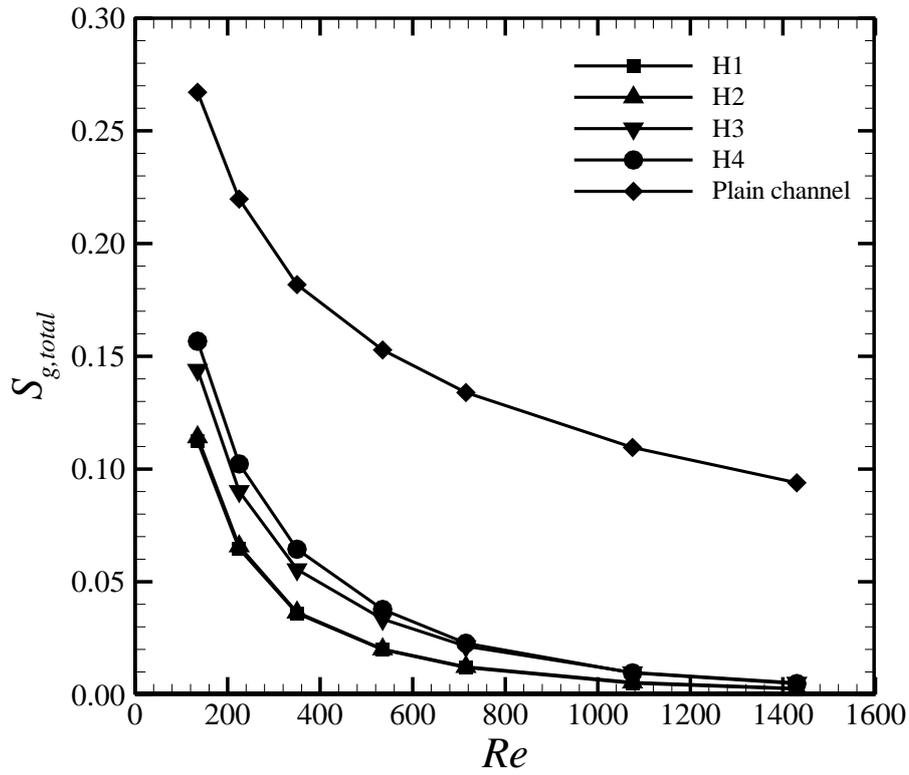

Figure 11- Dimensionless total entropy generation versus Reynolds number for different configurations. (a=w/2; $H_v$=w/4)

Table 1- Thermo-physical properties of pure water [16].

| Coolant | $k$ (W/mK) | $\mu$ (Pa.s) | $\rho$ (Kg/m3) | $c_p$ (J/KgK) |
|---------|------------|--------------|----------------|---------------|
| Pure Water | $0.6\times(1+0.00004167\times T)$ | $0.000002761\times\exp(1713/T)$ | 1000 | 4180 |

Table 2- The results of grid independency test for H1 configuration at $Re$ =715.

| Number of cells | $Nu_m$ | %Diff $Nu_m$ | $f$ | %Diff $f$ |
| --- | --- | --- | --- | --- |
| 787,400 (very coarse) | 25.3242 | -0.48 | 0.2586 | 3.45 |
| 952,400 (coarse) | 25.4155 | -0.12 | 0.2549 | 1.97 |
| 1,144,000 (intermediate) | 25.4949 | 0.19 | 0.2538 | 1.54 |
| 1,393,100 (fine) | 25.4524 | 0.02 | 0.2506 | 0.28 |
| 1,709,900 (very fine) | 25.4469 | - | 0.2499 | - |

Table 3- Comparison of pressure drop and heat transfer coefficient between numerical results and available experimental and numerical data.

| Pressure drop [Pa] | Re=465 | Re=933 | Re=1400 | Re=1865 |
|---|---|---|---|---|
| Experimental – Rectangular VGs [17] | 14.7 | 50.2 | 102.6 | 171.6 |
| Present study – Rectangular VGs | 13.8 | 48.4 | 101.2 | 169.4 |
| \|Difference (%)\| | 6.12 | 3.59 | 1.36 | 1.28 |
|  | Re=100 | Re=400 | Re=800 | Re=1600 |
| Numerical – Triangular VGs [17] | 1.1 | 4.8 | 14.5 | 51.1 |
| Present study – Triangular VGs | 1.2 | 5.0 | 15.0 | 52.2 |
| \|Difference (%)\| | 4.74 | 4.38 | 3.10 | 2.17 |
| Heat transfer coefficient [W.m$^{-2}$.K$^{-1}$] | Re=465 | Re=933 | Re=1400 | Re=1865 |
| Experimental – Rectangular VGs [17] | 342.9 | 465.5 | 513.9 | 554.7 |
| Present study – Rectangular VGs | 351.4 | 474.7 | 523.1 | 563.1 |
| \|Difference (%)\| | 2.48 | 1.98 | 1.79 | 1.51 |
|  | Re=100 | Re=400 | Re=800 | Re=1600 |
| Numerical – Triangular VGs [17] | 101.7 | 214.5 | 336.0 | 492.2 |
| Present study – Triangular VGs | 96.3 | 204.6 | 325.8 | 487.9 |
| \|Difference (%)\| | 5.61 | 4.84 | 3.13 | 0.88 |